\documentclass{aa}

\usepackage{graphicx}
\usepackage{txfonts}
\usepackage{color}
\usepackage{todonotes}

\begin{document}

   \title{Three dynamically distinct stellar populations in the halo of M49
   \thanks{Based on data collected at Subaru Telescope, which is operated by the National Astronomical Observatory of Japan under programme S14A-006 and with the William Herschel Telescope operated on the island of La Palma by the Isaac Newton Group of Telescopes in the Spanish Observatorio del Roque de los Muchachos of the Instituto de Astrof\'isica de Canarias.}}


\author{J. Hartke\inst{1} \and
        M. Arnaboldi\inst{1,2} \and
        O. Gerhard\inst{3} \and
        A. Agnello\inst{1} \and
        A. Longobardi\inst{4} \and
        L. Coccato\inst{1} \and
        C. Pulsoni\inst{3,5} \and
        K. C. Freeman\inst{6} \and
		    M. Merrifield\inst{7}
        }

\institute{European Southern Observatory,
        Karl-Schwarzschild-Str. 2, 85748 Garching, Germany\\
        \email{jhartke@eso.org}
        \and
        INAF, Observatory of Pino Torinese, Turin, Italy
        \and
        Max-Planck-Institut f\"{u}r Extraterrestrische Physik,
        Giessenbachstra{\ss}e, 85748 Garching, Germany
        \and
        Kavli Institute for Astronomy and Astrophysics,
        Peking University, 5 Yiheyuan Road, Haidian District, Beijing 100871, P. R. China
        \and
        Excellence Cluster Universe, Boltzmannstra{\ss}e 2, 85748, Garching, Germany
        \and
        RSAA, Mt. Stromlo Observatory, 2611 Canberra, Australia
        \and
		School of Physics and Astronomy, University of Nottingham, NG7 2RD, UK
		}
\date{\today}

\abstract
{M49 (NGC 4472) is the dominant galaxy in subcluster B of the Virgo Cluster, and
a benchmark for studying the build-up of the extended halos of brightest group
galaxies in the outskirts of galaxy clusters.}
{We investigate the kinematics in the outer halo of M49, look for substructures,
and describe the transition to the surrounding intra-group light.}
{As kinematic tracers we use planetary nebulae (PNe), combining kinematics from
the extended Planetary Nebula Spectrograph (PN.S) early-type galaxy survey
with our recent deep photometric sample. We study
the position-velocity-plane for bright and faint PN populations out to $95$ kpc
radius, and employ a multi-Gaussian model for the velocity distribution to
identify stellar populations with distinct kinematics and histories.}
{We report the detection of stellar-kinematic substructure associated with the
interaction of M49 with the dwarf irregular galaxy VCC 1249. We find two
kinematically distinct PN populations associated with the main M49 halo and the
extended intra-group light (IGL). These have velocity dispersions $\sigma_{\rm
halo}\simeq 170\;\mathrm{km}\;\mathrm{s}^{-1}$ and $\sigma_{\rm IGL}
\simeq400\;\mathrm{km}\;\mathrm{s}^{-1}$ at 10-80~kpc radii. The overall
luminosity profile and velocity dispersion at $\sim80$~kpc are consistent with a
flat circular velocity curve extrapolated from X-ray observations. The
dispersion of the PNe associated with the IGL joins onto that of the satellite
galaxies in subcluster B at $\sim100$~kpc radius. This is the first time that
the transition from halo to IGL is observed based on the velocities of
\emph{individual} stars.}
{Therefore the halo of M49, consisting of at least three distinct components,
has undergone an extended accretion history within its parent group potential.
The blue colours of the IGL component are consistent with a population of stars
formed in low-mass galaxies at redshift $\sim0.5$ that has since evolved
passively, as suggested by other data.}


   \maketitle
%

\section{Introduction}
\defcitealias{2013A&A...558A..42L}{L13}
\defcitealias{2015A&A...579L...3L}{L15b}
\defcitealias{2015A&A...579A.135L}{L15a}
\defcitealias{2017A&A...603A.104H}{H17}
\defcitealias{2017MNRAS.464..356V}{V17b}
\defcitealias{2003ApJ...591..850C}{C03}
\label{sec:intro}

In the current hierarchical paradigm, early-type galaxies (ETGs) grow with time. The initial phase of strong in-situ star formation is followed by growth through minor mergers and accretion at large radii \citep{1991ApJ...379...52W,2002NewA....7..155S,2010ApJ...725.2312O}. This scenario reproduces the morphology, metallicity, and the kinematics of the inner halos at $30-50$ kpc radius -- see, for example, evidence for recent accretion events at these radii in M87 \citep{2015A&A...579L...3L} and M49 \citep{2012A&A...543A.112A}.

As more effort is put into understanding the mass-assembly history, reaching  higher sensitivity and larger radii in these galaxies, recent extended photometry and detection of individual stars provide evidence for very extended, blue halos around nearby ETGs at the centre of their groups.  M49, the brightest galaxy in the Virgo Cluster, is surrounded by such an extended envelope that is characterised by a strong blueward colour gradient beyond two effective radii \citep{2013ApJ...764L..20M,  2017ApJ...834...16M}.
If this strong colour gradient, reaching from $B - V = 0.92$ at 25~kpc to $B - V = 0.65$ at 100~kpc, were due to a strong metallicity gradient, the corresponding stellar populations would have to have metallicities below [Fe/H] $\leq -0.2$ and $\leq -1$ respectively at these radii \citep{2013ApJ...764L..20M}\footnote{\citet{2013ApJ...764L..20M} assume \citet{2003MNRAS.344.1000B} single stellar population models for a range of ages. The values quoted here assume an age of at least 5~Gyr.}.

Within the cosmological framework, simulations predict that the mergers building up M49's halo would have typical mass ratios of $0.2\pm 0.1$ \citep[1:5,][]{2012ApJ...744...63O}.
However, according to the stellar mass-metallicity relation \citep{2017ApJ...847...18Z}, such merging satellites would deposit stars with much larger metallicities ([Fe/H] $\approx -0.1\pm0.1$), which would be redder in colour.
If the bulk of the metal-poor stars in the outer halo instead originated from the formation of many low-mass, low-metallicity objects at high redshift, this would be in tension with recent cosmological simulations \citep{2016MNRAS.458.2371R}, which predict that satellites with mass ratios less than $10^{-2} : 1$ do not contribute significantly to the mass of the stellar halo.
Finally, if the halo had been made by the more recent accretion of relatively massive (1:5 merger-mass ratio) star-forming galaxies, these would leave behind long-lasting dynamical signatures like asymmetric velocity fields or substructures in the velocity phase-space \citep{2005ApJ...635..931B}.

Therefore, a possible way to address the tension between the size and the colour of these halos at 100~kpc radial distance, and to discriminate between the different formation channels of these extended blue halos, is to measure the kinematics of the associated stars. If the halo formed from many small, metal-poor objects, its phase-space would be characteristically smooth and point symmetric. On the other hand, if it had been built relatively recently from more massive, star-forming satellites, we would detect dynamical signatures in the velocity phase-space.

While the dynamics of the halos of local group galaxies like M31 can be mapped by absorption-line spectroscopy of individual red-giant branch stars, at the distance of the Virgo Cluster these stars would have apparent magnitudes $m_V > 23.5$, which is fainter than the sky background. Integral-field spectrographs are ideally suited for mapping the central high-surface brightness regions of early-type galaxies \citep[ETGs, see e.g.][]{2002Ap&SS.281..363E,2014ApJ...795..158M,2016ARA&A..54..597C}, but due to their small field-of-view it becomes increasingly costly to map the extended halos at larger distances (beyond $2 r_\mathrm{e})$. Even with sparse sampling, the coverage does not extend beyond $4 r_\mathrm{e}$ \citep[e.g.][]{2009MNRAS.398..561W, 2014ApJ...795..158M, 2017MNRAS.471.4005B}.

Beyond these distances, discrete tracers like globular clusters (GCs) or planetary nebulae (PNe) are needed. The latter are the progeny of asymptotic giant branch (AGB) stars and can be easily identified due to their relatively strong [OIII]5007\AA\ emission lines, and the absence of a continuum. With multi-slit imaging techniques PNe can even be detected in galaxy clusters like Hydra~I \citep{2008AN....329.1057V} at 50~Mpc distance and Coma \citep{2005ApJ...621L..93G} at 100~Mpc distance. Within 20~Mpc distance, common techniques for PNe identification are the on-off band technique \citep{2002AJ....123..760A,2003AJ....125..514A} and counter-dispersed imaging \citep{2002PASP..114.1234D,2010A&A...518A..44M}.

Empirically, it has been found that PNe are about three times more abundant in the outer bluer halos around M87 and M49, compared to the inner redder galaxy \citep[][hereafter \citetalias{2013A&A...558A..42L,2015A&A...579A.135L,2017A&A...603A.104H}]{2013A&A...558A..42L,2015A&A...579A.135L,2017A&A...603A.104H}. This difference is also present in the morphology of the PN luminosity function (PNLF), as it has been found that PNLFs steepen from recent star-forming to old parent stellar populations \citep[e.g.][\citetalias{2015A&A...579A.135L, 2017A&A...603A.104H}]{2010PASA...27..149C}, thus making PNe ideal single stellar tracers to map the transition from the inner galaxy halo to the outer blue envelope in bright cluster galaxies.

In addition to PNe tracers being relatively abundant in blue outer halos, a series of studies indicate that PNe in different magnitude bins of the PNLF have distinct kinematics.
For example, the study of the kinematics of the PN population in NGC~4697, an ETG located in the Virgo southern extension, showed that PNe in different magnitude intervals of the PNLF exhibited different kinematics and spatial distributions \citep{2006AJ....131..837S}.
A very recent study of PN populations in the face-on disk galaxy NGC~628 showed that PNe within 0.5 mag from the bright cut-off of the PNLF are dynamically colder with respect to PNe at fainter magnitudes \citep[][Arnaboldi et al. 2018, in prep.]{2018MNRAS.476.1909A}. The correlation between the magnitude of the [OIII]5007\AA\ emission and kinematics may be understood in the framework given by the recent models of post-AGB evolution \citep{2016A&A...588A..25M}, which link the total luminosity of the PN core stars to the age and metallicity of the parent stellar population \citep{zijlstra_gesicki_bertolami_2016}.

In light of these empirical and theoretical results, we may expect that accretion events of satellites with different ages, masses, and metallicities, which build the extended halos of ETGs, will leave different imprints in the magnitude-position-velocity space traced by PN populations. Hence by studying correlations in this space, we may be able to identify substructures and assign them to different progenitors.

The aim of this work is to understand the build-up of M49's halo and the surrounding intra-group light by a synergy of accurate photometry \citepalias{2017A&A...603A.104H} from Suprime-Cam at Subaru Telescope \citep{2002PASJ...54..833M} with line-of-sight (LOS) velocities from the extended Planetary Nebula Spectrograph (ePN.S) ETG survey \citep{2017IAUS..323..279A,2017arXiv171205833P}.
The PN.S at the William Herschel Telescope is a double-arm slitless spectrograph that facilitates the identification of PNe and the measurement of their LOS velocities in a single observation \citep{2002PASP..114.1234D}.

This paper is organised as follows: in Sect.~\ref{sec:data} we describe the synergy of photometric and spectroscopic surveys of PNe in the halo of M49. Section~\ref{sec:methods} summarises the methods used for the analysis of the LOS velocity distribution (LOSVD).
We present our results in Sect.~\ref{sect:results} and subsequently discuss them in Sect.~\ref{sect:discussion}.
For the remainder of this paper we adopt a distance to M49 of $16.7$ Mpc \citep{2009ApJ...694..556B}, corresponding to a physical scale of $81$ pc per $1\arcsec$ and an effective radius of $r_\mathrm{e} = 194\arcsec.4\pm17\arcsec.0$ \citep[][corresponding to $15.7\pm1.4$~kpc]{2009ApJS..182..216K}.

\section{Matching photometric and kinematic data of PNe in M49}
\label{sec:data}
We aim to investigate whether there are correlations between substructures in distance-velocity phase-space with [OIII]5007\AA\ magnitudes for a large sample of PNe. We therefore combine accurate photometry from a narrow-band survey with LOS velocities obtained with the PN.S.
The complete photometric survey from \citetalias{2017A&A...603A.104H} consists of 624 PNe within a limiting magnitude of $m_\mathrm{5007,lim} = 28.8$, covering a major-axis distance of 155 kpc (equivalent to $9 r_{\mathrm{e}}$) from the centre of M49, excluding the central 13 kpc due to the high background from M49's light. For a detailed discussion of the survey's completeness and photometric accuracy we refer to \citetalias{2017A&A...603A.104H}.

The ePN.S ETG survey \citep{2017arXiv171205833P} provides positions (RA, dec), LOS velocities ($v$) and magnitudes ($m_{5007, \mathrm{PN.S}}$) for 465 PNe in the halo of M49, covering a major-axis distance of 95 kpc (equivalent to $6 r_{\mathrm{e}}$). We remove all measurements with a signal-to-noise ratio $S/N < 2.5$ in either of the two dispersed arms from the PN.S dataset. The median error on velocities measured with the PN.S is $\delta v = 20\;\mathrm{km}\;\mathrm{s}^{-1}$. We refer the reader to \citet{2007ApJ...664..257D} for a detailed description of the data reduction procedures for the PN.S.
We then calculate the velocity dispersion $\sigma = 363 \pm 17\;\mathrm{km}\;\mathrm{s}^{-1}$ using a robust-fitting technique \citep{2010A&A...518A..44M} and remove all PNe with velocities outside a velocity range of $3\sigma$ about the mean velocity $\bar{v} = 935 \pm 17\;\mathrm{km}\;\mathrm{s}^{-1}$.
We remove one high-velocity outlier that is likely not a PNe but a foreground object as its magnitude is much brighter than M49's bright cut-off $m^{\star}_{5007} = 26.8$, and 13 low-velocity outliers that are all located at the field edges and thus likely affected by degrading image quality. After the removal of low signal-to-noise measurements and $3\sigma$-clipping the cleaned catalogue contains 436 PNe.

From the cleaned catalogue, 226 PN.S-PNe fall into the unmasked regions of the photometric survey. We cross-correlate their positions with those of the PNe from the photometric survey with a tolerance of $5\arcsec$, resulting in 215 common objects. The resulting catalogue, containing positions (RA, dec), LOS velocities ($v$), and magnitudes ($m_{5007, \mathrm{Sub}}$) of 215 PNe, is referred to as \emph{matched catalogue}.
The combination of the two surveys allows us to assess the accuracy of the magnitudes measured with the PN.S in comparison to the excellent Suprime-Cam photometry. We find that an offset in magnitudes that is governed by a linear relation of the form
\begin{equation}
    m_\mathrm{5007,PN.S} = a m_\mathrm{5007,Sub} + b,
\end{equation}
with $a = 0.704\pm0.084$ and $b = 8.78 \pm 2.29$. We correct for the magnitude offset by inverting this relation. There is a residual $\pm 0.7\;\mathrm{mag}$ scatter of the PN.S magnitudes with respect to the Suprime-Cam photometry, which is likely due to instrumental effects as the PN.S has been optimised for kinematic measurements only. We apply the magnitude correction to all PN.S PNe in the cleaned catalogue, resulting in the \emph{magnitude-corrected PN.S catalogue}, which contains positions (RA, dec), LOS velocities ($v$), and magnitudes ($m_{5007, \mathrm{PN.S-corr}}$) of 441 PNe.

\section{Methods for LOSVD decomposition}
\label{sec:methods}
In this section, we briefly describe the methods used to analyse the closer velocity distribution (LOSVD) of PNe in the halo of M49.

\subsection{First and second moments of the LOSVD:}
Following the procedure outlined in \citet{2006AJ....131..837S}, we calculate the reduced velocity $U$ for each PNe with velocity $v$:
\begin{equation}
    U(x) = (v - v_\mathrm{sys})\cdot\mathrm{sgn}(x).
    \label{eq:U}
\end{equation}
M49's systemic velocity as computed from the ePN.S velocity fields is $v_\mathrm{sys} = 960\pm 24 \;\mathrm{km}\;\mathrm{s}^{-1}$ \citep{2017arXiv171205833P} and $x$ is the position along the galaxy's photometric major axis \citep[$PA = -31^{\degr}$,][]{2009ApJS..182..216K} .
PNe with $U > 0$ are denoted as co-rotating and with $U < 0$ as counter-rotating respectively.
The mean reduced velocity is calculated in magnitude bins to evaluate how the kinematics of M49 change as a function of magnitude.

We then calculate the LOS velocity dispersion profile as a function of radius, using the robust-fitting technique described in \citet{2010A&A...518A..44M}. As galaxies in group and clusters often have a LOSVD with strong wings due to the presence of intra-cluster or intra-group light \citep[IGL, e.g.][\citetalias{2015A&A...579A.135L}]{2012A&A...545A..37A,2018MNRAS.473.5446V,2018A&A...609A..78B}, we estimate the dispersion within three standard deviations from the mean velocity. The error on the velocity dispersion is calculated using Monte-Carlo techniques \citep[see][eq. (2)]{2010A&A...518A..44M}.

\subsection{Adaptive-kernel smoothing}
\label{ssec:smooth}
Smoothed velocity and velocity dispersion fields are calculated using the adaptive-kernel-smoothing technique of \citet{2009MNRAS.394.1249C}. For a sample of $N$ PNe, the smoothed velocity $\tilde{v}$ and $\tilde{\sigma}$ at every position on the sky ($x$, $y$) are
\begin{align}
	\tilde{v}(x, y) &= \frac{\sum_{i = 0}^N v_i w_i}{\sum_{i = 0}^N w_i} \\
	\tilde{\sigma}(x, y) &=  \left(\frac{\sum_{i = 0}^N v^2_i w_i}{\sum_{i = 0}^N w_i} - \tilde{v}(x,y)^2 - \delta v^2 \right)^{1/2},
\end{align}
where $v_i$ is the observed $i$th tracer velocity and $\delta v$ is the corresponding measurement error. Following \citet{2009MNRAS.394.1249C}, we use a distance-dependent Gaussian kernel for the $i$th tracer
\begin{equation}
	w_i = \exp \frac{-D_i^2}{2k(x, y)^2}
\end{equation}
where $D_i$ are the distances of the $i$th tracer from the remainder of the sample.
The kernel width $k$ is defined to be proportional to the distance $R_{i,M}$ to the $M$th closest tracers:
\begin{equation}
	k(x_i,y_i) = A R_{i,M}(x_M,y_M) + B,
\end{equation}
where $x_M$ and $y_M$ are the coordinates of the $M$th closest PN to the $i$th tracer $(x_i, y_i)$.
\citet{2017arXiv171205833P}
optimised the kernel parameters $A$ and $B$ for the best compromise between spatial resolution and noise smoothing through Monte-Carlo simulations of discrete velocity fields extracted from cosmological simulations (see Sect 3.1.1 therein). For M49, they determine $A = 0.73$ and $B = 11.3$ as kernel parameters.
We use these parameters to calculate the two-dimensional smoothed velocity and velocity dispersion fields presented in Sect.~\ref{sect:results}.
\subsection{A multi-Gaussian model for LOSVD decomposition}
\label{ssec:methods_model}
The information of the velocity and velocity dispersion fields is contained in the LOSVD. We model the LOSVD as the sum of individual Gaussians to evaluate whether it is consistent with a single or multiple components.
In order to assign PNe to the kinematic components, we use a Gaussian-mixture method \citep[see e.g.][\citetalias{2015A&A...579L...3L}]{2011ApJ...742...20W,2012MNRAS.419..184A,2013MNRAS.436.2598W,2014MNRAS.442.3299A}. We decompose the LOSVD profiles into multiple independent Gaussian components. Each Gaussian $G_i$ can be described by its central value $\mu_i$ and dispersion $\sigma_i$. For a PN $k$ with measured velocity $v_k\pm\delta v_k$ this is
\begin{equation}
    G_i(v_k, \delta v_k; \mu_i,\sigma_i) = \frac{1}{\sqrt{2\pi}\sigma_i}\exp\left(\frac{(v_k - \mu_i)^2}{2(\sigma_i^2 + \delta v_k^2)}\right).
\end{equation}
In the simple case of a two-component system, the likelihood would be
\begin{equation}
    \mathcal{L}_k = \frac{G_1(v_k, \delta v_k; \mu_1,\sigma_1) + a_2/a_1 G_2(v_k, \delta v_k; \mu_2,\sigma_2)}{1 + a_2/a_1},
\end{equation}
with $a_1$ and $a_2$ being the amplitudes of the respective Gaussians. The total likelihood is the product of the individual likelihoods that are calculated for each \emph{individual} PN.
This method allows us to exploit the information of every PN without the need for binning.

We assume flat priors on the model parameters and calculate their posterior probability distribution functions (PDFs) using the ensemble-based MCMC sampler \textsc{emcee} \citep{2013PASP..125..306F}.

\section{Results}
\label{sect:results}
\subsection{Distinct kinematics of bright and faint PNe}
\begin{figure}
\centering
   \includegraphics[width=8.8cm]{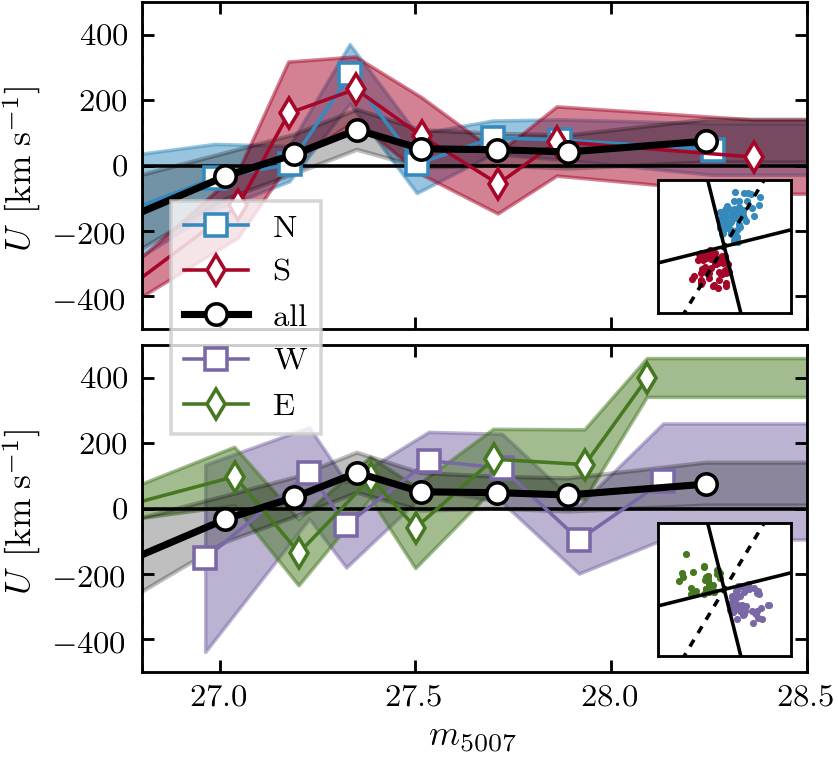}
     \caption{Mean reduced velocity $U$ versus magnitude $m_{5007}$ of PNe in the matched catalogue. \textit{Top:} Northern (blue squares) and Southern (red diamonds) quadrants. \textit{Bottom:} Western (purple squares) and Eastern (green diamonds) quadrants. In both panels, the black open symbols denote the entire sample and the respective colour bands the $1\sigma$-error on the reduced velocity. The insets highlight the respective quadrants on the sky, with M49's photometric major axis denoted by a dashed line. The excess of bright PNe with positive reduced velocity in the Northern and Southern quadrants motivates the division into a bright and a faint PN sample.}
     \label{fig:U}
\end{figure}

\begin{figure*}
\sidecaption
   \includegraphics[width=12cm]{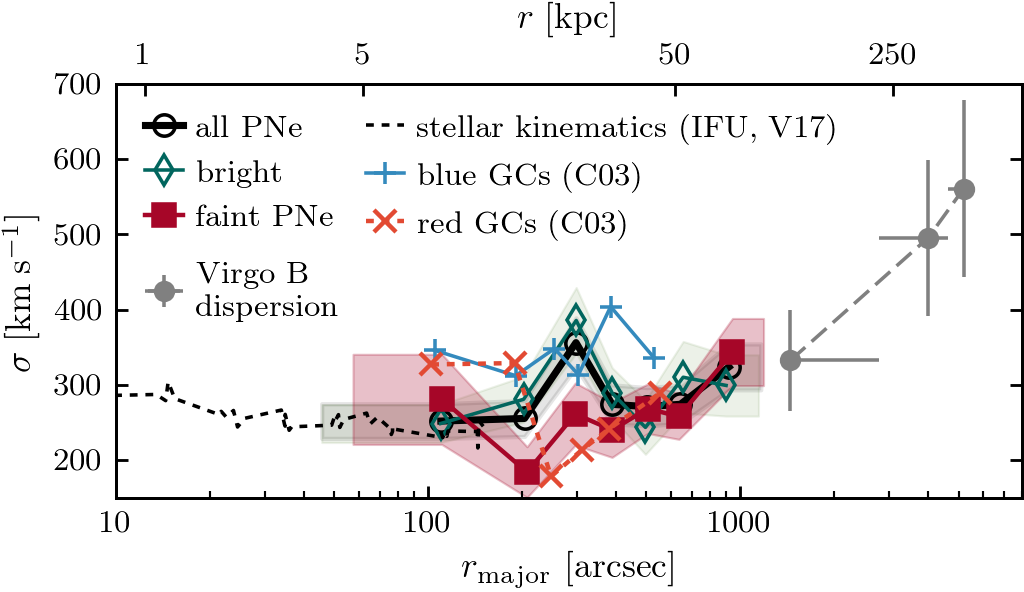}
     \caption{LOS velocity dispersion profile as a function of major-axis radius of the total (open black circles), bright (open green diamonds), and faint (filled red squares) magnitude-corrected PNe in the halo of M49. The coloured bands denote the $1\sigma$ errors. The stellar velocity dispersion profile is indicated by the dashed black line \citepalias{2017MNRAS.464..356V} and the dispersions of metal-poor and metal-rich GCs are indicated by the solid blue and dashed red line respectively \citepalias{2003ApJ...591..850C}. The grey error bars connected with dashes denote the velocity dispersion of galaxies in the Virgo Subcluster B. The velocity dispersion of the faint PN-sample reaches that of the Subcluster B at large radii, which indicates that these PNe are tracing the underlying intra-group light.}
     \label{fig:sigma}
\end{figure*}

We want to investigate whether the kinematics of PNe correlate with magnitude as possible correlations or anti-correlations could point to different stellar populations with distinct kinematics. We therefore calculate the reduced velocity for PNe from the matched catalogue using eq.~\eqref{eq:U}, as these have the most accurate magnitude measurements. As Fig.~\ref{fig:U} illustrates, we find no significant correlation between reduced velocity and magnitude, however, the entire sample (black circles) shows an excess of co-rotating PNe brighter than $m_{5007} < 27.5\;\mathrm{mag}$.
This excess is driven by PNe localised in the northern (blue squares) and southern (red diamonds) quadrants, as illustrated in the top panel of Fig.~\ref{fig:U}. The PNe in the western (purple squares) and eastern (green diamonds) quadrants do not show such an excess; see bottom panel in Fig.~\ref{fig:U}. A two-dimensional Kolmogorov-Smirnov (K-S) test \citep{1983MNRAS.202..615P,1987MNRAS.225..155F} investigating whether the northern and southern samples could be drawn from the same $U$-magnitude distributions as the western and eastern samples results in a $p$-value of $p = 0.17$, thus rejecting this hypothesis \citep{Press:2007:NRE:1403886}.
Based on the behaviour of the mean reduced velocity as a function of magnitude we divide the magnitude-corrected PN.S-sample into a \emph{bright} ($m_{5007} < 27.5$, 258 PNe) sample that is governed by an excess reduced velocity in the fields along the major axis, and a \emph{faint} ($m_{5007} > 27.5$, 178 PNe) sample, where this excess is absent.

The comparison of the LOS velocity dispersion profile, calculated in seven elliptical bins, of the magnitude-corrected sample with that of stellar tracers from integral-field spectroscopy \citep[][hereafter \citetalias{2017MNRAS.464..356V}]{2017MNRAS.464..356V} shows that the PN kinematics agree well with those of the integrated light in the region of overlap, see solid and dashed black lines in Fig.~\ref{fig:sigma}. We then calculate the velocity dispersion profiles for the bright (open green diamonds) and faint (filled red squares) samples separately.

\begin{figure}[h!]
\begin{center}
\includegraphics[width=8.8cm]{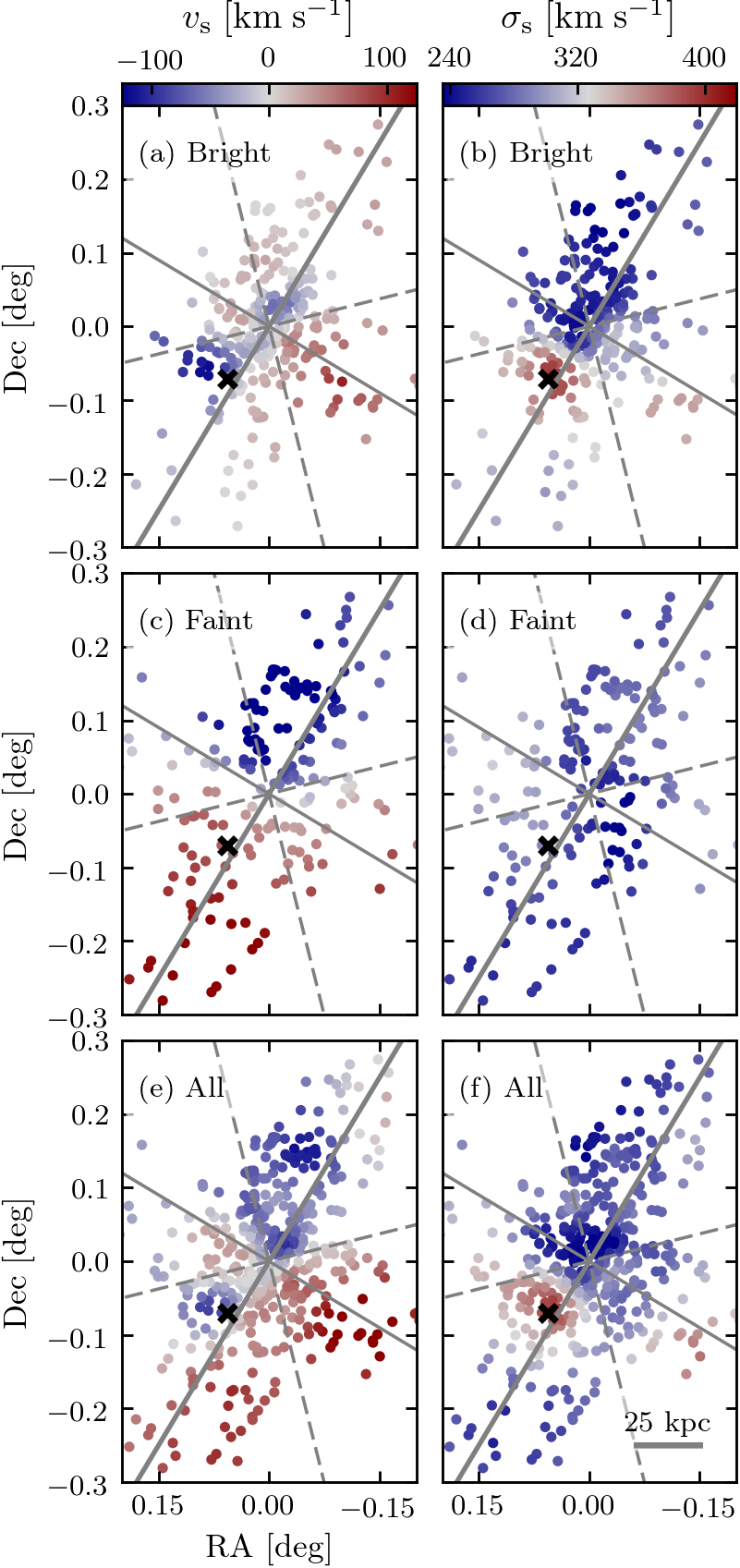}
  \caption{Smoothed velocity (left column) and velocity dispersion (right column) fields for the bright (top row) and faint (middle row) magnitude-corrected PN-samples. For comparison, the bottom row shows the smoothed fields for the total sample from \citet{2017arXiv171205833P}. The solid grey lines denote the photometric major and minor axes of M49. The dashed lines divide the sample into four quadrants. The black cross denotes the position of the dwarf irregular galaxy VCC 1249.
  The mean errors on the smoothed fields are
  $\Delta v_\mathrm{bright} = 39\;\mathrm{km}\;\mathrm{s}^{-1}$,
  $\Delta v_\mathrm{faint} =  43 \;\mathrm{km}\;\mathrm{s}^{-1}$,
  $\Delta \sigma_\mathrm{bright} = 34 \;\mathrm{km}\;\mathrm{s}^{-1}$, and
  $\Delta \sigma_\mathrm{faint} = 38  \;\mathrm{km}\;\mathrm{s}^{-1}$ respectively.
  The bar in the lower-left corner of panel (f) corresponds to 25~kpc.
  North is up, east is to the left.}
  \label{fig:vel_fields}
\end{center}
\end{figure}

We also construct smoothed velocity and velocity dispersion fields for the total PN-sample as well as for the bright and faint samples, using the adaptive kernel-smoothing technique \citep{2009MNRAS.394.1249C,2017arXiv171205833P} that is described in Sect.~\ref{ssec:smooth}. The resulting fields are shown in Fig.~\ref{fig:vel_fields}, where the smoothed values $\tilde{v}$ and $\tilde{\sigma}$ are plotted at the position of the observed PNe. The bright and faint samples have distinct features which we will discuss in turn. It is apparent that the velocity and velocity dispersion of the bright sample that are shown in panels (a) and (b) of Fig~\ref{fig:vel_fields} differ significantly from those of the faint sample shown in panels (c) and (d).

\subsection{Faint sample: separation of halo and intra-group light}
\begin{figure*}
\begin{center}
	\includegraphics[width=18cm]{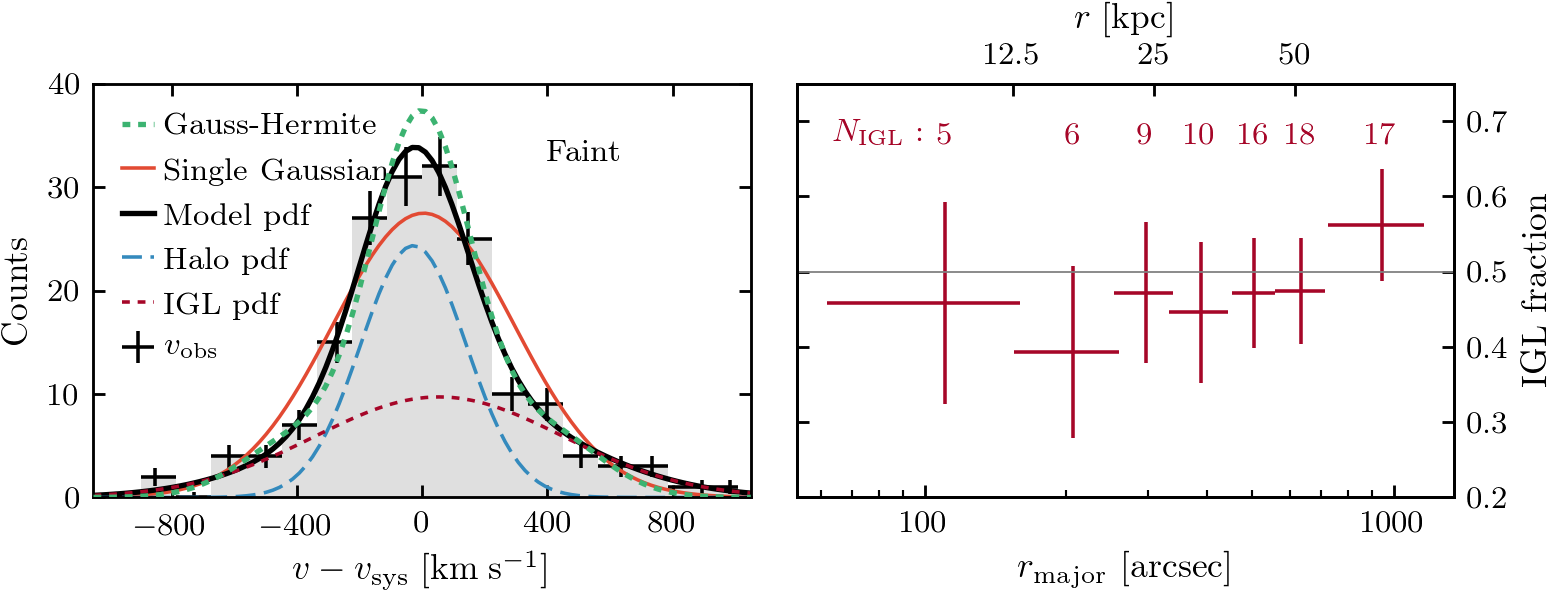}
  \caption{\textit{Left:} LOSVD of the faint component (grey histogram with errorbars) and best-fit model pdf (black line), consisting of the halo (dashed blue line) and IGL (dotted red line), and scaled to the observed LOSVD. While the likelihood is fit with a bin-free method, we bin the LOSVD for visual convenience. For comparison, the best-fit single-Gaussian model is shown in orange. The dashed green line denotes the best-fit Gauss-Hermite expansion. \textit{Right:} Fraction of faint PNe associated with the IGL as a function of major-axis radius, based on the probabilities derived from the LOSVD decomposition. The number of PNe associated with the IGL is given on top of each bin.}
  \label{fig:LOSVD_faint}
\end{center}
\end{figure*}

\begin{table}
\caption{Maximum likelihood and Bayesian Information Criteria (BIC) of single- and double-Gaussian models of the LOSVD in the respective PN-samples. $N_\mathrm{Gauss}$ denotes the number of Gaussians and $N_\mathrm{fit}$ the number of fitted parameters.}              
\label{tab:bic}      
\centering                                      
\begin{tabular}{c c c r c }          
\hline\hline                        
Sample & $N_\mathrm{Gauss}$ & $N_\mathrm{fit}$ &  $\log(\mathcal{L}_\mathrm{max})$ & BIC \\    
\hline                                   
Faint & 1 & 2 &  $-1352$ & 2714 \\      
Faint & 2 & 5 & $-1258$ & 2543 \\
\hline
Bright-North & 1 & 2 & $-834$ & 1678\\
Bright-North & 2 & 5 & $-835$ & 1694 \\
\hline
Bright-South & 1 & 2 & $-1009$ & 2028 \\
Bright-South & 2 & 2 & $-1005$ & 2020 \\
\hline                                             
\end{tabular}
\end{table}

The velocity field of the faint sample (Fig.~\ref{fig:vel_fields}c) is dominated by a rotation of $v_\mathrm{rot} = 120\;\mathrm{km}\;\mathrm{s}^{-1}$ about the minor axis. The velocity dispersion (Fig.~\ref{fig:vel_fields}d) is $\sigma = 270\;\mathrm{km}\;\mathrm{s}^{-1}$ along the major axis and increases up to $\sigma = 320\;\mathrm{km}\;\mathrm{s}^{-1}$
along the minor axis. Comparing the spatial density of tracers, the faint component (see Fig.~\ref{fig:vel_fields}c) is less centrally concentrated compared to the bright one (Fig.~\ref{fig:vel_fields}a), which allows us to study the kinematics of the extended outer halo.

The smoothed velocity (Fig.~\ref{fig:vel_fields}c) and velocity dispersion (Fig.~\ref{fig:vel_fields}d) fields do not immediately suggest multiple velocity components due to their smooth and symmetric appearance. However, inspecting the LOS velocity dispersion profile shown with red squares in Fig.~\ref{fig:sigma}, the rising dispersion at large radii -- even reaching the velocity dispersion of the Virgo Subcluster B \citep[grey error bars, derived from][]{1993A&AS...98..275B} -- indicates a smooth transition that might be driven by an additional component with larger velocity dispersion.

We therefore constructed the LOSVD for the PNe in the faint sample. As shown by the histogram in the left panel of Fig.~\ref{fig:LOSVD_faint} it has strong wings.
If we characterise the overall LOSVD of the faint component in terms of Gauss-Hermite polynomials \citep{1993ApJ...407..525V,1993MNRAS.265..213G} with moments $\mu, \sigma, h_3$, and $h_4$, we obtain a high $h_4$ value, i.e. $h_4 = 0.11\pm0.03$ and a much smaller $h_3 = 0.01\pm0.03$ (dashed green line in Fig.~\ref{fig:LOSVD_faint}).
This is in contrast to observations of ETGs in equilibrium, where the asymmetric deviations of the LOSVD from Gaussianity (characterised by $h_3$) are more prominent than the symmetric deviations (characterised by $h_4$), with the latter being a few percent at most \citep{1993MNRAS.265..213G,1994MNRAS.269..785B}.

\begin{figure}
\begin{center}
	\includegraphics[width=8.8cm]{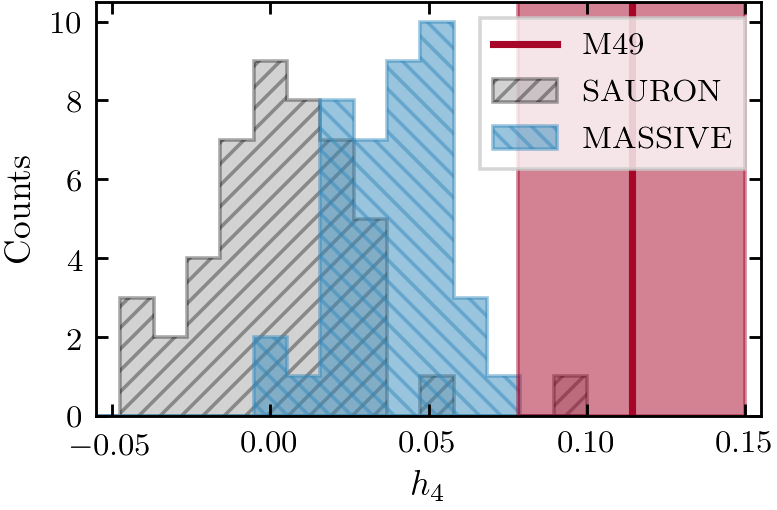}
  \caption{Best-fit $h_4$ value and error band (red vertical line and shaded region) derived from the LOSVD of the faint sample compared to the $h_4$ distribution of 47 and 41 ETGs from the SAURON \citep{2004MNRAS.352..721E} and MASSIVE \citep{2017MNRAS.464..356V} surveys (grey and blue histograms).}
  \label{fig:h4}
\end{center}
\end{figure}

In order to illustrate the uncommonly large value of $h_4$ with respect to galaxies in equilibrium, we compare our derived value to the distribution of luminosity-averaged $h_4$ values of ETGs in the SAURON \citep{2004MNRAS.352..721E} and MASSIVE \citep{2017MNRAS.464..356V} surveys. As shown in Fig.~\ref{fig:h4}, the error-band of our best-fit value of $h_4$ (indicated by the red shaded region) overlaps with the tails of the histograms from the SAURON (grey) and MASSIVE (blue) surveys and is larger than any of the measured $h_4$ values based on IFU data. It shows that the radially extended LOSVD of the faint PN sample cannot be reproduced by the LOSVDs found in most ETGs within a few effective radii ($0.5 - 4.0\,r_\mathrm{e}$). The large $h_4$ value measured for M49 is more extreme than any of the 41 values measured for the MASSIVE galaxies, which demonstrates that it has not been drawn from the same distribution at the $97.5\%$~confidence level.

We note that M49 is part of the MASSIVE sample and that the $h_4$ value measured here is larger than the luminosity-averaged value determined by \citet{2017MNRAS.464..356V}. However, our best-fit value agrees with their $h_4$ value at the largest radii within the errors \citep[see][Fig~D8]{2017MNRAS.464..356V}, continuing the trend of rising $h_4$ values with increasing radius observed in the MASSIVE survey.
This positive gradient of $h_4$ values with radius present in the MASSIVE survey is also illustrated by the offset between the mean $h_4$ values between the SAURON and MASSIVE surveys. The SAURON survey covers the central $0.5-1\,r_\mathrm{e}$, where the $h_4$ values are generally small or even slightly negative \citep[see also][]{1994MNRAS.269..785B}. The radial coverage of the MASSIVE survey extends to $2.0-4.0\,r_\mathrm{e}$. Due to this larger coverage, the transition to a core-wing structure of the LOSVD as signalled by an increase in $h_4$ value can be observed. Our measurement that includes data out to $6\,r_\mathrm{e}$ shows that this positive gradient of $h_4$ values extends to larger radii than those sampled in the MASSIVE survey.

Strongly winged LOSVDs as the one observed in M49 are also observed in other galaxies in group and cluster environments \citep[e.g.][\citetalias{2015A&A...579A.135L}]{2011A&A...528A..24V,2018MNRAS.473.5446V,2018A&A...609A..78B}.
Hydrodynamical cosmological simulations of galaxies at the centre of clusters find that the strongly-winged LOSVDs of star particles are well described by two Gaussians with different dispersion \citep{2014MNRAS.437..816C,2010MNRAS.405.1544D} and centred on different mean velocities \citep{2005MNRAS.361.1203V}.  The Gaussian with the largest velocity dispersion indicates the presence of a population of intra-group (or intra-cluster) stars at large radii, which do not trace the halo but the group (or cluster) potential.

Using the method described in Sect.~\ref{ssec:methods_model}, we fit a single- and a double-Gaussian model to the data, with the latter being clearly favoured, with a lower Bayesian Information Criterion (BIC, cf. Table~\ref{tab:bic}).
If we were to fit a single Gaussian to the LOSVD for the entire faint sample its velocity dispersion would be $\sigma_\mathrm{faint,single} = 290^{+17}_{-16}\;\mathrm{km}\;\mathrm{s}^{-1}$ and it would be centred on M49's systemic velocity. A K-S test whether the faint LOSVD can be drawn from a single Gaussian with these parameters results in $p = 0.08$. This model is denoted by the orange line in the left panel of Fig.~4, while the best-fit two-component model scaled to the binned LOSVD is denoted by the solid black line. For comparison, we also carry out a K-S test between the observed faint LOSVD and our best-fit two-component model, resulting in $p = 0.64$.

The first component of the model (dashed blue line), which we associate with M49's halo, is centred on $\mu_\mathrm{halo} = -27.1^{+26.8}_{-24.5}\;\mathrm{km}\;\mathrm{s}^{-1}$ (with respect to the systemic velocity of M49) and has a velocity dispersion of $\sigma_\mathrm{halo} = 169 \pm 27 \;\mathrm{km}\;\mathrm{s}^{-1}$.
The second component (dotted red line), which we associate with the IGL, has a central velocity of $\mu_\mathrm{IGL} = 54.0^{+54.0}_{-55.0}\;\mathrm{km}\;\mathrm{s}^{-1}$ that is marginally offset from the halo one and its velocity dispersion $\sigma_\mathrm{IGL} = 397^{+38}_{-36} \;\mathrm{km}\;\mathrm{s}^{-1}$, which is more than twice as large as that of the halo component, or, phrased differently, is $5.9\sigma$ away.
The difference between the central velocities of the IGL and main halo indicates a velocity bias that is consistent with that observed in several cD galaxies \citep[e.g.][\citetalias{2015A&A...579A.135L}]{2005MNRAS.361.1203V,2018A&A...609A..78B}.

In order to quantify the contribution of the second Gaussian with respect to the main one, we statistically associate each PN to either one of the two Gaussians based on its LOS velocity. To determine the mean fraction and its error as a function of radius, we calculate $10\,000$ realisations. On average, we associate 81 PNe with the second, broader Gaussian. These PNe have a uniform spatial distribution in M49's halo. The right panel of Fig.~\ref{fig:LOSVD_faint} shows the increasing relative contribution of PNe associated to the second Gaussian as a function of major-axis radius.
The fraction of PNe associated to the second Gaussian, which is already $30\%$ in the innermost elliptical bin, rises with increasing radius and reaches above $50\%$ beyond 50~kpc.

\subsection{Bright sample: kinematic signature of the accretion of VCC 1249}

We now investigate the presence of a kinematically distinct component in the bright sample. North of the minor axis, the velocity dispersion is $\sigma = 230\;\mathrm{km}\;\mathrm{s}^{-1}$ and the velocity field is featureless, with the majority of the PNe moving with velocities close to M49's systemic velocity (see Fig.~\ref{fig:vel_fields}a).
The velocity field south of the minor axis shows a multi-peaked distribution. There is an approaching component $5\arcmin$ along the major axis with a $\sigma = 400\;\mathrm{km}\;\mathrm{s}^{-1}$ peak in the velocity dispersion field. This peak is centred on the position of the dwarf irregular galaxy VCC 1249 (black cross), which is currently interacting with M49 \citep[e.g.][]{2012A&A...543A.112A}.
VCC 1249 has a systemic velocity of $v_\mathrm{VCC1249,sys}=390\pm30\;\mathrm{km}\;\mathrm{s}^{-1}$ (SDSS), hence the superposition of the LOS velocities of PNe originating from this galaxy and halo PNe results in a high velocity dispersion in the kernel smoothing \citep[recall $v_\mathrm{M49,sys} = 960\pm 24 \;\mathrm{km}\;\mathrm{s}^{-1}$,][]{2017arXiv171205833P}.

We therefore divide the bright population along the major axis into two subsamples. The southern subsample contains signatures of the accretion of VCC 1249. Based on the featureless appearance of the smoothed velocity and velocity dispersion fields north of the minor axis (see Fig.~\ref{fig:vel_fields}), we assume that the PN-kinematics in the northern subsample are unaffected by the accretion of VCC 1249. We again fit single- and double-Gaussian models to the data and find that the best-fit model for the bright northern subsample is a single Gaussian with $\mu = -2.7^{+25.1}_{-24.8}\;\mathrm{km}\;\mathrm{s}^{-1}$ and $\sigma = 270^{+19}_{-51}\;\mathrm{km}\;\mathrm{s}^{-1}$. Adding a second Gaussian to the model only marginally increases the maximum likelihood and is penalised with a higher BIC (see Table~\ref{tab:bic}). A K-S test remains inconclusive whether the northern subsample is drawn from a single or a double Gaussian model.

\begin{figure}[!t]
	\includegraphics[width=8.8cm]{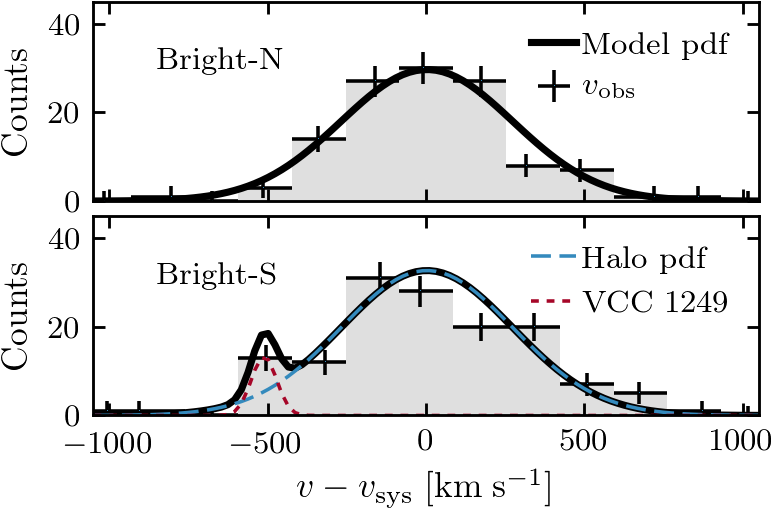}
  \caption{LOSVD of the bright component (gray histogram with errorbars) in the north (top) and south (bottom) of the galaxy. The northern component is best-fit by a single Gaussian (solid black line). The substructure of VCC 1249 in the southern part is fit by a Gaussian (dotted red line) in addition to the halo (dashed blue line). While the likelihood is fit with a bin-free method, we bin the LOSVDs for visual convenience.}
  \label{fig:LOSVD_bright}
\end{figure}

Assuming that the underlying halo population of bright PNe in M49 is symmetric, we expect the LOSVD of halo PNe\footnote{PNe not associated to the accretion of VCC 1249} in the southern subsample to have the same mean velocity and velocity dispersion that we fit in the northern subsample. However, when comparing the LOSVD of PNe in the northern subsample with that of the southern sample (gray histograms in the top and bottom panel of Fig.~\ref{fig:LOSVD_bright}), it is apparent that the LOSVD of the southern subsample cannot be described by a single Gaussian. This is further corroborated by \citet{2017arXiv171205833P}, who cannot reconcile the asymmetric LOSVD of M49 at the position of VCC 1249 with simulated galaxies in equilibrium.

\begin{figure}
	\includegraphics[width=8.8cm]{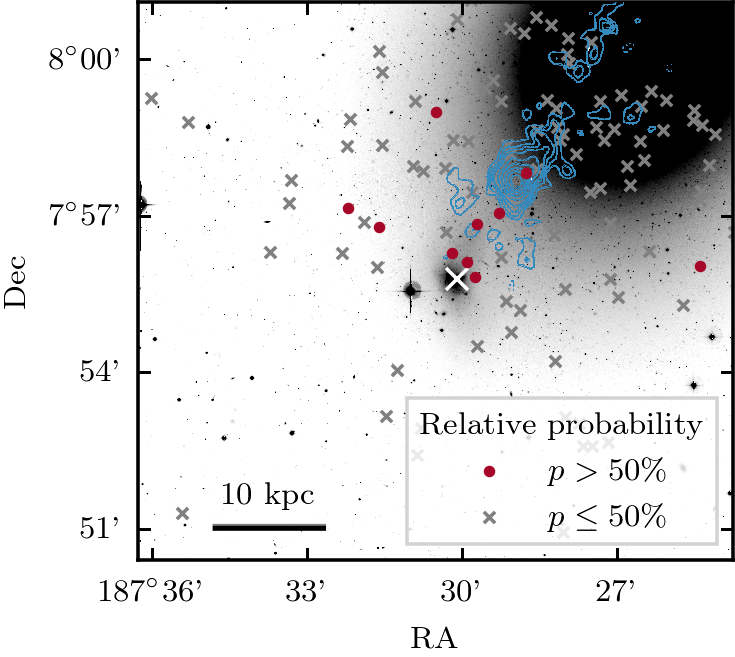}
  \caption{NGVS $g$-band image of M49 \citep[from][]{2012A&A...543A.112A}. PNe from the southern bright PN sample are colour-coded by their relative probability to be associated with VCC 1249. The white cross denotes the centre of VCC 1249. The \ion{H}{i} contours from \citet[][precessed to from B1950 to J2000]{1994AJ....108..844M} are superposed in blue. The bar in the lower-right corner corresponds to 10~kpc. North is up, east is to the left.}
  \label{fig:VCC}
\end{figure}

In order to account for the contribution of PNe from the accretion of VCC 1249, we fit a double-Gaussian model: the first Gaussian is fixed by the Gaussian fit to the LOSVD of PNe from the northern sample and the second Gaussian is free to vary. We find that the second component is centred on $\mu_\mathrm{VCC1249} = -512\pm30\;\mathrm{km}\;\mathrm{s}^{-1}$ relative to the systemic velocity of M49 and has a velocity dispersion of $\sigma_\mathrm{VCC1249} = 40^{+28}_{-19} \;\mathrm{km}\;\mathrm{s}^{-1}$. The maximum likelihood and BIC for this model in comparison to a single Gaussian are shown in Table~\ref{tab:bic}, with the two-component model being preferred with a lower BIC.
If we add M49's systemic velocity to the centre of the secondary component, it results in $v_\mathrm{VCC1249,LOS} = 412\pm30\;\mathrm{km}\;\mathrm{s}^{-1}$, which agrees with VCC 1249's systemic velocity from SDSS within the errors. As illustrated in Fig.~\ref{fig:VCC}, PNe associated with this secondary peak cluster around the position of VCC 1249, which further supports the association of these PNe with VCC 1249.

\section{Discussion and conclusions}
\label{sect:discussion}

In this work, we found that the PNe in the halo of M49 divide in subpopulations in a position-velocity-magnitude space. The three subpopulations correspond to the main halo, the intra-group light, and the accreted satellite VCC 1249.
We investigated the bright and the faint samples and identified the main halo in both. In the faint sample, we find evidence for two kinematic components: the M49 halo and the IGL with velocity dispersions $\sigma_\mathrm{faint,M49} = 169 \pm 27 \;\mathrm{km}\;\mathrm{s}^{-1}$ and $\sigma_\mathrm{faint,IGL} = 397^{+38}_{-36} \;\mathrm{km}\;\mathrm{s}^{-1}$, respectively. If only a single Gaussian is fit, the resulting velocity dispersion is $\sigma_\mathrm{faint,single} = 298^{+17}_{-16}\;\mathrm{km}\;\mathrm{s}^{-1}$.
In the bright component, where a single Gaussian fit is preferred based on a lower BIC, the velocity dispersion is $\sigma_\mathrm{bright} = 270^{+19}_{-51}\;\mathrm{km}\;\mathrm{s}^{-1}$, which is in agreement with the estimate for the faint component, obtained with the same method, within the errors. Therefore the absence of two distinct IGL and halo model components in the bright sample is likely due to the smaller sample size.

The fit of a two-component LOSVD for the bright sample, with contribution from a high-dispersion IGL component and a halo with velocity dispersion similar to the one measured in the faint component, is impeded by the kinematic signature of the dwarf irregular galaxy VCC 1249 in the low-velocity tail of the LOSVD.
We therefore first discuss the accretion of VCC 1249 in Sect.~\ref{ssect:vcc1249} and then the implications on IGL detection in the bright component in Sect.~\ref{ssect:IGLdisc}.

\subsection{Accretion of VCC 1249}
\label{ssect:vcc1249}
We can clearly detect the kinematic signature of the dwarf irregular galaxy VCC 1249 in the bright PN sample. The interaction between M49 and VCC 1249 led to the formation of stellar tidal tails and to the removal of gas from the dwarf \citep{2012A&A...543A.112A}.
Our best-fit velocity for the VCC 1249 PNe ($412\pm30\;\mathrm{km}\;\mathrm{s}^{-1}$) falls in between the systemic velocity derived from SDSS $v_\mathrm{sys} = 390 \pm 30\;\mathrm{km}\;\mathrm{s}^{-1}$ and the peak velocity of the \ion{H}{i} removed from it, which is $v_\mathrm{HI} = 469\;\mathrm{km}\;\mathrm{s}^{-1}$
\citep{1994AJ....108..844M}, suggesting that the PNe trace both the dwarf and the stripped stars as their average velocity is between the \ion{H}{I} and VCC 1249 systemic velocities.
PNe associated with the interaction of VCC1249 and M49 are clustered around the position of the dwarf irregular galaxy with a loose association of PNe appearing to stream towards or from the centre of M49 (see Fig.~\ref{fig:VCC}). They broadly overlap in space with the \ion{H}{i} tail associated with VCC 1249 \citep[][blue contours]{1994AJ....108..844M}. The interaction also triggered the formation of \ion{H}{ii} regions, which also emit at $5007$\AA. However, as they are extended they should not be contaminating our PNe sample. A cross-correlation of our catalogue with the positions of \ion{H}{ii} regions from \citet{2012A&A...543A.112A} did not yield any matches. Thus our detections are not \ion{H}{ii} regions, but PNe associated with this accretion event. Our findings will be the basis for a more in-depth dynamical model of the dynamics of M49's halo and its recent interaction with VCC 1249.

\subsection{M49 halo and the IGL in the Virgo Subcluster B}
\label{ssect:IGLdisc}
We find evidence for an extended population of PNe that are kinematically associated with the IGL.
Based on their LOS velocities, we statistically associate 81 PNe in the faint sample with the IGL population, which are uniformly distributed in M49's halo.
This PN subsample has a velocity dispersion more than twice that of the halo ($\sigma_\mathrm{IGL} = 397^{+38}_{-36} \;\mathrm{km}\;\mathrm{s}^{-1}$) and traces the IGL out to 95 kpc along the major axis, much further than any X-ray observations of M49 to date.
It is therefore difficult to assess the dynamical state of the IGL based on the X-ray potential.
If we however assumed that the circular velocity profiles derived by \citet{2010MNRAS.409.1362D} remained flat beyond their maximum radius of 30~kpc, the circular velocity would be $v_\mathrm{c} \approx 500 \;\mathrm{km}\;\mathrm{s}^{-1}$.
\citet{2010MNRAS.404.1165C} derive a number of relations to calculate the circular velocity from the velocity dispersion for a given surface brightness profile and orbital configuration.
Inverting their eq. (24) for an isotropic system, assuming a flat rotation curve, and the surface brightness profile from \citet{2009ApJS..182..216K}, the corresponding dispersion at 80~kpc would thus be $\sigma \approx 300\;\mathrm{km}\;\mathrm{s}^{-1}$, similar to the combined velocity dispersion of the halo and IGL traced by the PNe at large radii (see Fig.~\ref{fig:sigma}). This suggests that the circular velocity inferred from X-ray observations within 30~kpc is indeed constant out to $\sim80$~kpc, and that the transition to the group structure takes place at larger radii, as also shown in Fig.~\ref{fig:sigma}.

In order to trace the transition from halo to IGL, we consider the velocity dispersion profile.
In the decomposition of the LOSVD of the faint sample, we approximate the velocity dispersion profile of the IGL as constant, resulting in with a value of $\sigma_\mathrm{IGL} = 397^{+38}_{-36} \;\mathrm{km}\;\mathrm{s}^{-1}$.
However, the velocity dispersion profile of the total faint sample (indicated by purple squares in Fig.~\ref{fig:sigma}) increases with radius, reaching the velocity dispersion of galaxies in the Subcluster B (grey error bars).
The subcluster velocity dispersion was calculated from the velocities of individual galaxies \citep{2014ApJS..215...22K} within $1.^{\degr}6$ from the subcluster centre based on their membership classification by \citet{1993A&AS...98..275B}.
The characteristic increase in velocity dispersion is also observed in the bright sample.
A similar increase in velocity dispersion is seen in the metal-rich (red) GCs, however, as their data have a smaller radial extent, \citet{2003ApJ...591..850C} did not attribute this rise to an IGL contribution.
The right panel of Fig.~\ref{fig:LOSVD_faint} illustrates that the rise in velocity dispersion is due to the increased fraction of PNe associated with the IGL at larger radii.

The fact that the velocity dispersion of the faint sample reaches that of the Subcluster B indicates that M49's outermost halo is dynamically controlled by the subcluster's gravity. This is one of the defining features of a cD, or in this case group dominant galaxy. Early attempts to measure the velocity dispersion out to radii where the halo is dynamically controlled by cluster gravity were made by \citet{1979ApJ...231..659D}. In the case of NGC 6166 in Abell 2199, where a velocity dispersion profile reaching the cluster dispersion was already observed by \citet{2002ApJ...576..720K}, the cD halo would have not been recognisable based of surface photometry alone \citep{2015ApJ...807...56B}. \emph{M49 is the first galaxy where the transition from galaxy to cluster velocity dispersion has been measured from the velocities of individual stars (PNe).}

In the case of M49, deep broad-band photometry from the VEGAS survey \citep{2015A&A...581A..10C,2017A&A...603A..38S} and \citet{2013ApJ...764L..20M,2017ApJ...834...16M} already indicated the presence of an extended, blue outer halo, that is characterised by a steep colour gradient and a shallower surface-brightness profile.
Based on their narrow-band PN survey, \citetalias{2017A&A...603A.104H} argue for the presence of IGL, that is traced by a PN population with a PN-specific frequency of $\alpha_\mathrm{IGL} = (1.03\pm0.22)\times10^{-8}\;\mathrm{PN}\;L^{-1}_{\sun,\mathrm{bol}}$, three times higher than the specific frequency of M49's halo, and an underlying stellar population that contributes 10\% of the total luminosity in the surveyed area. The PN kinematics presented here corroborate their result of an extended PN population associated with the IGL. The blueward colour gradient measured by \citet{2013ApJ...764L..20M} constrain the IGL colour to $B-V \leq 0.65$.

In addition to the smooth blue halo, \citet{2010ApJ...715..972J} observe a number of shells surrounding M49. However, these shells are significantly redder than the underlying smooth blue halo, and thus clearly of distinct origin. \citet{2013ApJ...764L..20M} argue that the colour of these shells ($B-V \approx 0.85$) are in agreement with the accretion of a single dwarf elliptical (dE) galaxy from the dE population surrounding the Virgo Cluster, which have a mean colour of $B-V \approx 0.77$ \citep{2004AJ....128.2797V}.

\subsection{Origin of the IGL in the Virgo Subcluster B}
In addition to the photometric properties discussed in the previous subsection, we would like to derive constraints on the dynamical age of the IGL, on the basis of its smooth appearance in photometry \citep{2013ApJ...764L..20M,2015A&A...581A..10C} and regular velocity field and LOSVDs (see Fig.~\ref{fig:vel_fields} and Fig.~\ref{fig:LOSVD_faint}). The smooth appearance suggests that the IGL has a dynamical age that is older than the orbital precession time, i.e. at least 5~Gyr. Hence it rules out the accretion of a moderately massive (1:5 merger-mass ratio), young star-forming galaxy as the origin of the extremely blue IGL.
As the LOSVD is smooth and the velocity field is point-symmetric, this is consistent with the accretion of many low-mass systems. Due to the constraint on the dynamical age ($>5$~Gyr) and the observed $B-V$ colour ($0.6$), these systems have to have metallicities lower than $\mathrm{[Fe/H]} < -1 \pm 0.5$ \citep{2013ApJ...764L..20M}. According to the low-mass end of the mass-metallicity relation \citep{2013ApJ...779..102K}, galaxies with these low metallicities would have stellar masses of at most few $10^8\;M_{\sun}$.

Independent constraints on the age and metallicity of intracluster stellar populations in the Virgo Cluster were derived from HST-ACS observations of $\sim5300$ stars in a single pointing located between M87, M86, and M84 at $190$~kpc distance from M87 \citep{2007ApJ...656..756W}. The theoretical stellar evolution models suggest that this intracluster population is dominated by stars with old ages ($\geq 10$~Gyr) with a significant number of very metal-poor stars ($\mathrm{[Fe/H]} < -1.5$). If the IGL is as blue as the last measured point at 100~kpc by \citet[][$B-V\approx0.6$]{2013ApJ...764L..20M}, for a 10-Gyr old stellar population the corresponding metallicity is $\mathrm{[Fe/H]} < -1.5$, leading to even stronger constraints on the stellar masses of the satellites that built the IGL:
these would have to be less than $M_{\star} \leq 10^7\;M_{\sun}$ \citep{2013ApJ...779..102K}.

We now put these measurements into the context of galaxy clusters observed at slightly larger redshift ($0.3 \leq z < 0.6$).
Recent studies on six clusters in the Hubble Frontier Fields \citep{2017ApJ...837...97L} have measured negative colour gradients with increasing distance from the brightest cluster galaxy (BCG) out to the intra-cluster light \citep[ICL,][]{2017arXiv171003240M,2017ApJ...846..139M,2018MNRAS.474.3009D}. The three different studies agree that these strong colour gradients cannot be explained with metallicity gradients alone, but are a combination of metallicity and age gradients.

\citet{2017ApJ...846..139M} extend these measurements out to 300~kpc from the central BCG and measure colours as blue as $B-V \approx 0.1$. They estimate that $10-15\%$ of the ICL mass beyond 150~kpc was contributed by (post-) star-forming populations with young ages ($\sim1$~Gyr) at $z \approx 0.5$ and low stellar mass ($M_{\star} < 3\times10^{9}\;M_{\sun}$).
However, the total ICL mass in the six Hubble Frontier Field clusters is similar to the mass of all existing red satellites with stellar masses below $10^{10}\;M_{\sun}$ as estimated from the combination of the Frontier Field data with the GLASS survey
\citep{2017ApJ...835..254M}, which suggests that tidally stripped stars from higher-mass systems likely also contribute some of the mass of the ICL \citep{2017ApJ...846..139M}. Due to dynamical friction, low-mass systems will stay roughly at the radius at which they were disrupted, while the more massive (and thus metal-rich and red) satellites would sink towards the BCG centre \citep[e.g.][]{1975ApJ...202L.113O,2017MNRAS.464.2882A}.

An evolutionary path for the formation of a blue, smooth, and extended IGL around M49 may come from the early (before $z \sim 0.5$) accretion of a number of low-mass satellites with  young ($\sim1$~Gyr) and blue ($B-V = 0.1$) stellar populations at that redshift. Since their accretion more than 5~Gyr ago, their stars would have evolved passively and now reached a colour of $B-V = 0.65$ at $z=0$. If applicable, the constraints on the age from the resolved stellar populations in the ICL in Virgo \citep{2007ApJ...656..756W} would set the time of accretion at even higher redshift and the stellar mass of the satellites to lower masses (e.g. $10^{7}\;M_{\sun}$)

The formation of a smooth IGL by the accretion of low-mass satellites at intermediate redshift can be well reproduced in N-body simulations combined with particle tagging, e.g. \citet{2017MNRAS.464.2882A}, where it is assumed that any low-mass halo forms stars according to the cosmological baryon fraction.
The contribution of such low-mass systems with low merger-mass ratios is lower in current hydrodynamical cosmological simulations: while being able to resolve mergers with mass ratios below $10^{-4}:1$, the cumulative effect from mergers with stellar mass ratios below $10^{-2}:1$ is found to be negligible with respect to stellar halo growth \citep[see Fig.~1 in][]{2016MNRAS.458.2371R}.
As more massive satellites form sufficient stars in these hydrodynamical simulations, these satellites can now well reproduce the build-up of the halo closer to M49. See for example \citet{2017Galax...5...34P}, who reproduce shells that are distinctly redder than the underlying galaxy halo, like the ones observed in M49 \citep{2010ApJ...715..972J,2013ApJ...764L..20M}.

Our observations suggest that the IGL has formed early from the accretion of many low-mass satellites. This scenario is also in line with the observations of GC colours in ETGs; blue, metal-poor GCs are attributed to have been accreted from satellite galaxies while red, metal-rich GCs to have been formed in-situ \citep{2003ApJ...591..850C,2006ApJ...639...95P,2014ApJ...796...10L,2017MNRAS.465.3622R,2016MNRAS.457.1242F}.
Our strong results on the mass of the satellites ($M_{\star} \lesssim 10^8\;M_{\sun}$), which were accreted at redshift $z \approx 0.5$ implies that the feedback in the current cosmological simulations is too strong for low-mass satellites to form sufficient stars. As these satellites need to contribute significantly to the growth of the IGL, the feedback would have to be reduced early in their lifetimes.

Understanding the build-up of the IGL in low-mass subclusters is important because they will eventually contribute to the overall ICL in their parent cluster. Our results corroborate the existence of multiple channels contributing to the ICL and IGL build-up at different cluster mass scales and distances to the cluster centre. In order to constrain the masses of the IGL and halo progenitor galaxies, we are planning to obtain more accurate LOS velocity measurements, as they would allow us to detect also substructures with lower velocity dispersion and hence lower masses, which will set important constraints on the feedback mechanisms in the current cosmological simulations.

\section*{Acknowledgements}
We greatly acknowledge the support and advice of the ING and Subaru staff.
We thank Fabrizio Arrigoni Battaia for the provision of the coordinates of the \ion{H}{ii} regions and the NGVS $g$-band image of M 49.
JH thanks I. S\"{o}ldner-Rembold and C. Spiniello for useful comments and discussions. JH and CP furthermore acknowledge support from the IMPRS on Astrophysics at the LMU Munich.
This research made use of \textsc{astropy} \citep{2013A&A...558A..33A}, \textsc{corner} \citep{ForemanMackey2016}, \textsc{emcee} \citep{2013PASP..125..306F}, \textsc{matplotlib} \citep{Hunter:2007}, and \textsc{numpy} \citep{2011arXiv1102.1523V}.
This research has made use of the SIMBAD database, operated at CDS, Strasbourg, France and the Sloan Digital Sky Survey. Funding for the Sloan Digital Sky Survey IV has been provided by the Alfred P. Sloan Foundation, the U.S. Department of Energy Office of Science, and the Participating Institutions. The authors thank the anonymous referee for the constructive comments and the careful reading of the manuscript.

\bibliographystyle{aa} 
\bibliography{M49_letter}
\end{document}